\documentstyle[aps,12pt]{revtex}

\begin{document}
\draft
\title{Efficient microwave-induced optical frequency conversion}
\author{D.V. Kosachiov\thanks{%
permanent address: Tyumentransgas Co, 627720 Yugorsk, Russia} and E.A.
Korsunsky}
\address{Institut f\"{u}r Experimentalphysik, Technische Universit\"{a}t Graz, A-8010
Graz, Austria}
\date{\today{}}
\maketitle

\begin{abstract}
Frequency conversion process is studied in a medium of atoms with a $\Lambda
$ configuration of levels, where transition between two lower states is
driven by a microwave field. In this system, conversion efficiency can be
very high by virtue of the effect of electromagnetically induced
transparency (EIT). Depending on intensity of the microwave field, two
regimes of EIT are realized: ''dark-state'' EIT for the weak field, and
Autler-Townes-type EIT for the strong one. We study both cases via
analytical and numerical solution and find optimum conditions for the
conversion.
\end{abstract}


Frequency conversion is a useful technique for generation of coherent
tunable radiation \cite{boyd}. Efficient conversion of a continuous-wave
(c.w.) radiation at relatively low pump intensities requires high nonlinear
optical susceptibility of an (atomic) medium, which can be achieved by
tuning to resonances. However, this will also increase the medium absorption
and refraction seriously limiting the conversion efficiency. It was recently
proposed \cite{harr90} and demonstrated in many experiments that this
problem can be overcome if one uses the effect of electromagnetically
induced transparency (EIT) \cite{harr97}. For example, the UV radiation has
been generated by use of dc electric-field coupling in atomic hydrogen \cite
{hak91}. Radiation fields have been used to produce transparency in
experiments on red to blue frequency conversion with molecular sodium \cite
{bab96}, and on enhanced four-wave mixing with doped crystals \cite{hemm97}.
Recently, blue to UV \cite{jain96} and UV to VUV \cite{mer99} conversion in
atomic Pb vapor have been reached with almost unity photon-conversion
efficiency.

EIT is due to quantum interference in multilevel quantum systems (atoms,
molecules, dopants in solids) induced by applied electromagnetic radiation.
There are two basic mechanisms responsible for EIT. The first one occurs at
large strength of one, ''coupling'', electromagnetic field which mixes and
splits quantum states (Autler-Townes effect). When another, weaker ''probe''
field is tuned in between the two mixed states, it experiences no absorption
not only because of the splitting but also due to interference between
excitation paths to two mixed states. This mechanism works well even in the
case when the states mixed by the coupling field decay spontaneously. The
second mechanism takes place at comparable intensities of applied fields. In
this case, the cancellation of absorption and refraction can be explained by
creation of a coherent superposition of atomic states (''dark'' state) not
excited by the radiation, and by preparation of atoms in this superposition
(which is termed coherent population trapping - CPT) \cite{cptrev}. The dark
state should be stable in order to allow for the population trapping.
Therefore, the dark state should be a superposition of the metastable atomic
states.

In the present paper we consider the frequency conversion in a scheme where
both mechanisms of EIT are possible. This is a three-level $\Lambda $ system
(Fig. 1), where $\left| 1\right\rangle -\left| 3\right\rangle $ and $\left|
2\right\rangle -\left| 3\right\rangle $ are the dipole-allowed optical
transitions, and the microwave (m.w.) transition $\left| 1\right\rangle
-\left| 2\right\rangle $ is a magnetic-dipole one. Such systems can be
realized, e.g., on D-lines in alkali atoms, and may also be found in some
molecules and doped crystals as well. Experiments on the absorption
reduction induced by the m.w. field have recently been performed with
similar systems in solids \cite{zha97,ham98,yam98}. The scheme is
interesting, above all, because it allows easy control of the frequency
conversion process by intensity of the microwave field. Possible
applications of the present system include generation of the optical field
which is phase- and amplitude-correlated to the input field, optical phase
conjugation \cite{hem95}, generation of squeezed light \cite{kol93}, as well
as quantum noise suppression and quantum correlation \cite{hem97,luk99}.

When one of the optical fields (let say, $\omega _{31}$) and the m.w. field $%
\omega _{m}$ are applied to the $\Lambda $ atom, they induce an optical
susceptibility on transition $\left| 2\right\rangle -\left| 3\right\rangle $%
. This leads to the generation of the optical field with frequency $\omega
_{32}$. In general, this field as well as the field $\omega _{31}$ will be
quickly absorbed if they are tuned close to the resonance. However, the
absorption can be substantially reduced for particular values of the
microwave intensity. For the weak m.w. field, optical waves create the dark
state which is only slightly disturbed. The strong m.w. field mixes and
splits both ground states $\left| 1\right\rangle $ and $\left|
2\right\rangle $ so that relatively weak optical fields experience EIT of
the Autler-Townes type. Here we consider both cases, treating an interaction
of the e.m. radiation with atoms as well as the propagation of radiation
through the medium in exact manner.

The optical waves propagation along the $z-$axis in the medium is described
by the Maxwell equations. In the slowly varying amplitude and phase
approximation, and in continuous-wave limit these equations can be reduced
to the following form \cite{kos95,kor99}:
\begin{mathletters}
\label{Maxw5}
\begin{eqnarray}
\frac{dg_{n}}{d\zeta } &=&-%
\mathop{\rm Im}%
(\tilde{\sigma}_{3n}), \\
\frac{d\varphi _{n}}{d\zeta } &=&-\frac{1}{g_{n}}%
\mathop{\rm Re}%
(\tilde{\sigma}_{3n}),
\end{eqnarray}
where $g_{n}=d_{3n}E_{n}/2\hbar \gamma _{31}$ are the dimensionless optical
field amplitudes (Rabi frequencies), $E_{n}$ and $\varphi _{n}$ $\left(
n=1,2\right) $ are the optical amplitudes and phases, respectively; $\gamma
_{31}$ is the spontaneous decay rate in a channel $\left| 3\right\rangle
\rightarrow \left| 1\right\rangle $, and $d_{3n}=\langle 3|{\bf \hat{d}}%
|n\rangle $ are the matrix elements of the electric-dipole moment operator $%
{\bf \hat{d}}${\bf \ }in the basis of atomic states $|l\rangle ,l=1,2,3$.
The dimensionless optical length $\zeta $ is expressed in terms of an
absorption cross-section for the optical field $\zeta =\left( 2\pi \omega
_{_{31}}d_{31}^{2}/c\hbar \gamma _{31}\right) Nz=\left( 3\pi c^{2}/2\omega
_{_{31}}^{2}\right) Nz$, $N$ is the density of active atoms.

The medium optical polarization components (the right-hand side of Eqs. (\ref
{Maxw5})) are determined by the (steady-state) density matrix elements $%
\sigma _{3n}$ averaged over the atomic velocities with the distribution $%
w\left( v_{z}\right) $, where $v_{z}$ is the $z$-projection of the atom
velocity: $\tilde{\sigma}_{3n}=\int\limits_{-\infty }^{+\infty
}dv_{z}w\left( v_{z}\right) \sigma _{3n}\left( v_{z}\right) $, with $\sigma
_{3n}\left( v_{z}\right) =\rho _{3n}\left( v_{z}\right) \exp \left[ i\left(
\omega _{3n}t-k_{3n}z+\chi _{3n}\right) \right] $, where $\rho _{3n}\equiv
\langle 3|\hat{\rho}|n\rangle $, ${\hat{\rho}}$ is the atomic density
matrix. The phase $\chi _{3n}$ is the sum of the e.m. field phase $\varphi
_{3n}$ and the phase $\vartheta _{3n}$ of the atomic dipole moment $%
d_{3n}=\left| d_{3n}\right| {\rm e}^{i\vartheta _{3n}}$: $\chi _{3n}=\varphi
_{3n}+\vartheta _{3n}$. Similar quantities are determined for the microwave
transition: Rabi frequency $g_{m}=\mu H/2\hbar \gamma _{31}$ with the m.w.
field amplitude $H$ and phase $\varphi _{m}$, and matrix element$\,\mu
\equiv \langle 1|{\bf \hat{\mu}}|2\rangle $ of the magnetic-dipole moment $%
{\bf \hat{\mu}}$.

Presence of the field on $\left| 1\right\rangle -\left| 2\right\rangle $
transition and/or both optical fields means, corresponding to the Maxwell
equations, that the m.w. wave should also change along the propagation path.
The EIT-assisted generation of a microwave radiation has recently been
observed in atomic Cs vapor \cite{god98}. We, however, will not consider
this effect here since the changes are of the order of $\Delta
g_{m}^{2}\approx \left( \omega _{_{m}}/\omega _{_{31}}\right)
g_{1}^{2}\approx \,\left( 10^{-8}\div 10^{-5}\right) g_{1}^{2}$ at the most
\cite{kor00} which is negligible in the present context. Moreover, the
propagation direction of the m.w. wave (traveling or standing one in a m.w.
cavity) can be chosen perpendicular to the $z-$axis.

Let us now consider the case when all three e.m. fields are in resonance
with corresponding transitions. This situation can be studied analytically
if we additionally suppose equal spontaneous relaxation rates $\gamma _{31}=$
$\gamma _{32}\equiv \gamma $, zero relaxation rate of the coherence between
states $\left| 1\right\rangle $ and $\left| 2\right\rangle $: $\Gamma =0$,
and zero atomic velocity $v_{z}=0$. Solution of the density-matrix equations
for this case is given in the earlier work of one of us \cite{kos95}.
Nevertheless, we display it here again since it is important for further
consideration:
\end{mathletters}
\begin{mathletters}
\label{ims}
\begin{eqnarray}
\mathop{\rm Im}%
(\sigma _{31}) &=&-\frac{g_{2}g_{m}g_{0}^{2}\left(
g_{0}^{2}-2g_{m}^{2}\right) }{2L}\sin \Phi +\frac{g_{m}^{2}g_{1}\left(
g_{1}^{2}-g_{2}^{2}+2g_{2}^{2}\sin ^{2}\Phi \right) }{L}, \\
\mathop{\rm Im}%
(\sigma _{32}) &=&\frac{g_{1}g_{m}g_{0}^{2}\left(
g_{0}^{2}-2g_{m}^{2}\right) }{2L}\sin \Phi -\frac{g_{m}^{2}g_{2}\left(
g_{1}^{2}-g_{2}^{2}-2g_{1}^{2}\sin ^{2}\Phi \right) }{L},
\end{eqnarray}
\end{mathletters}
\begin{mathletters}
\label{res}
\begin{eqnarray}
\mathop{\rm Re}%
(\sigma _{31}) &=&\frac{g_{2}g_{m}\left( g_{1}^{2}-g_{2}^{2}\right) \left(
g_{0}^{2}-2g_{m}^{2}\right) }{2L}\cos \Phi +\frac{g_{m}^{2}g_{1}g_{2}^{2}}{L}%
\sin 2\Phi , \\
\mathop{\rm Re}%
(\sigma _{32}) &=&-\frac{g_{1}g_{m}\left( g_{1}^{2}-g_{2}^{2}\right) \left(
g_{0}^{2}-2g_{m}^{2}\right) }{2L}\cos \Phi -\frac{g_{m}^{2}g_{1}^{2}g_{2}}{L}%
\sin 2\Phi ,
\end{eqnarray}
and the excited state population is given by
\end{mathletters}
\begin{equation}
\rho _{33}=\frac{g_{m}^{2}\left[ \left( g_{1}^{2}-g_{2}^{2}\right)
^{2}+4g_{1}^{2}g_{2}^{2}\sin ^{2}\Phi \right] }{L},  \label{ro3}
\end{equation}
with
\begin{eqnarray*}
L &=&\frac{1}{2}g_{0}^{6}+g_{m}^{2}\left[ 3\left( g_{1}^{2}-g_{2}^{2}\right)
-2g_{0}^{4}+2g_{0}^{2}+12g_{1}^{2}g_{2}^{2}\sin ^{2}\Phi \right]
+2g_{0}^{2}g_{m}^{4}, \\
g_{0}^{2} &=&g_{1}^{2}+g_{2}^{2},
\end{eqnarray*}
and the relative phase $\Phi $ is determined as
\begin{equation}
\Phi =\left( \chi _{31}-\chi _{32}\right) -\chi _{12},  \label{Phi}
\end{equation}
where $\chi _{12}=\varphi _{m}+\vartheta _{12}$ ( $\mu =\left| \mu \right|
{\rm e}^{i\vartheta _{12}}$) is the phase of the m.w. transition.

One sees from Eqs. (\ref{ims}-\ref{ro3}), that the medium is absolutely
transparent and not refractive for
\begin{equation}
\Phi =\pi n,\,\,n=0,1,2,...  \label{phi-cond}
\end{equation}
and
\begin{equation}
g_{1}=g_{2}.  \label{g-cond}
\end{equation}
These are exactly the conditions for the dark state in closed $\Lambda $
system \cite{kos95,kos92,buc86,kos91}.

For arbitrary optical field amplitudes and phases, however, the refraction
and absorption (or amplification) of individual frequency components can be
substantial. Here we are interested in generation of the optical field $%
\omega _{32}$ with lowest possible losses of the total e.m. energy. The
change of the total energy flow is proportional to intensity $%
I=I_{31}+I_{32} $ of the optical waves. The intensity is expressed in terms
of Rabi frequency as $I_{n}=$ $\left( c/8\pi \right) E_{n}^{2}=$ $\left(
2\hbar \omega _{3n}^{3}/3\pi c^{2}\right) g_{n}^{2}\gamma $, so that $%
dI/dz\sim \left( dg_{1}^{2}/d\zeta +dg_{2}^{2}/d\zeta \right) \,=-2\left(
g_{1}%
\mathop{\rm Im}%
\left( \sigma _{31}\right) +g_{2}%
\mathop{\rm Im}%
\left( \sigma _{32}\right) \right) =-2\rho _{33}$, where the last equality
follows from the steady-state density matrix equations \cite{kos95}. Thus,
we arrive at almost obvious conclusion that the dissipation of the e.m.
energy is small when the excited state population is small: $\rho _{33}\ll 1$%
. Analysis of the expression (\ref{ro3}) shows that, for arbitrary $%
g_{1},\,g_{2}$ and $\Phi $, this is the case for two ranges of the m.w. Rabi
frequency: $g_{m}\ll 1,\,g_{0}$ and $g_{m}\gg 1,\,g_{0}$. These values
correspond to EIT of the CPT-type and the Autler-Townes-type, respectively.

The change of the fields can be calculated analytically in present situation
\cite{arm62}. An interesting feature of the resonant case is that the phase
equation can be solved for arbitrary values of $g_{m}$. The propagation
equation for the relative phase is as follows (if we neglect the change of
the m.w. field phase $\varphi _{m}$ and the atomic dipole phases $\vartheta
_{ns}$ along the propagation path):
\[
\frac{d\Phi }{d\zeta }=-\left( \frac{1}{g_{1}}%
\mathop{\rm Re}%
(\sigma _{31})-\frac{1}{g_{2}}%
\mathop{\rm Re}%
(\sigma _{32})\right) .
\]
One can obtain from Eqs. (\ref{ims}) and (\ref{res}) that $\frac{1}{g_{1}}%
\mathop{\rm Re}%
(\sigma _{31})-\frac{1}{g_{2}}%
\mathop{\rm Re}%
(\sigma _{32})=\left( \cos \Phi /\sin \Phi \right) \left( g_{2}%
\mathop{\rm Im}%
(\sigma _{31})+g_{1}%
\mathop{\rm Im}%
(\sigma _{32})\right) /g_{1}g_{2}$ so that
\begin{equation}
\frac{d\Phi }{d\zeta }=-\frac{\cos \Phi }{\sin \Phi }\frac{1}{g_{1}g_{2}}%
\left( g_{2}%
\mathop{\rm Im}%
(\sigma _{31})+g_{1}%
\mathop{\rm Im}%
(\sigma _{32})\right) =\frac{\cos \Phi }{\sin \Phi }\frac{1}{g_{1}g_{2}}%
\frac{d\left( g_{1}g_{2}\right) }{d\zeta }.  \label{Mph}
\end{equation}
which can immediately be integrated to give the constant of motion:
\begin{equation}
g_{1}g_{2}\cos \Phi =\Pi .  \label{Cph}
\end{equation}
The constant $\Pi $ is determined from the boundary conditions at the $\zeta
=0$. In particular, when one optical field is generated, $g_{2}(\zeta =0)=0$%
, we have constant value of $\cos \Phi $:
\begin{equation}
\cos \Phi (\zeta )=0.  \label{cos0}
\end{equation}

We now consider both EIT cases separately. For a weak m.w. field, $g_{m}\ll 1
$, the density matrix elements to the second order in $g_{m}$ are:
\begin{mathletters}
\label{ims-1}
\begin{eqnarray}
\mathop{\rm Im}%
(\sigma _{31}) &=&-\frac{g_{2}g_{m}}{g_{0}^{2}}\sin \Phi +\frac{%
2g_{m}^{2}g_{1}\left( g_{1}^{2}-g_{2}^{2}+2g_{2}^{2}\sin ^{2}\Phi \right) }{%
g_{0}^{6}}, \\
\mathop{\rm Im}%
(\sigma _{32}) &=&\frac{g_{1}g_{m}}{g_{0}^{2}}\sin \Phi -\frac{%
2g_{m}^{2}g_{2}\left( g_{1}^{2}-g_{2}^{2}-2g_{1}^{2}\sin ^{2}\Phi \right) }{%
g_{0}^{6}}.
\end{eqnarray}
For generation of the field $\omega _{32}$ the dissipation of total optical
energy is proportional to (taking into account Eq. (\ref{cos0})):
\end{mathletters}
\[
\frac{dg_{0}^{2}}{d\zeta }=-\frac{4g_{m}^{2}}{g_{0}^{2}}\,,
\]
which has a solution
\begin{equation}
g_{0}^{4}=g_{0}^{4}(\zeta =0)-8g_{m}^{2}\zeta .  \label{diss1}
\end{equation}
At sufficiently small optical length $\zeta \ll 1/8g_{m}^{2}$, the total
intensity decays linearly: $g_{0}^{2}=g_{0}^{2}(\zeta =0)-\left(
4g_{m}^{2}/g_{0}^{2}(\zeta =0)\right) \zeta $. If we neglect this slow decay
(which would simply correspond to neglect of terms of the second order in $%
g_{m}$), we obtain the following amplitude equations
\begin{eqnarray*}
\frac{dg_{1}}{d\zeta } &=&-\frac{g_{2}g_{m}}{g_{0}^{2}} \\
g_{2}^{2} &=&g_{0}^{2}-g_{1}^{2}
\end{eqnarray*}
which can be easily solved:
\begin{mathletters}
\label{sol1}
\begin{eqnarray}
g_{1}^{2} &=&g_{0}^{2}\cos ^{2}\left( \frac{g_{m}}{g_{0}^{2}}\zeta \right) ,
\\
g_{2}^{2} &=&g_{0}^{2}\sin ^{2}\left( \frac{g_{m}}{g_{0}^{2}}\zeta \right) .
\end{eqnarray}
The solution indicates that the e.m. energy is transferred back and forth
between two optical waves as the optical length increases. The period of
these oscillations is $\zeta _{\pi }=\pi g_{0}^{2}/g_{m}$, which is much
smaller than the characteristic length of the total energy dissipation: $%
\zeta _{diss}\approx g_{0}^{2}(\zeta =0)/4g_{m}^{2}$, cf. Eq. (\ref{diss1}).
Therefore, very efficient conversion takes place at
\end{mathletters}
\begin{equation}
\zeta _{\max }=\pi g_{0}^{2}(\zeta =0)/2g_{m}.  \label{zmax1}
\end{equation}
The loss of the optical intensity is $\Delta I\,/\,I=2\pi g_{m}\ll 1$ at
this point.

The reason for such an efficient process is a preparation of the medium in
almost dark state. If the e.m. field between the states $\left|
1\right\rangle $ and $\left| 2\right\rangle $ is not applied then the dark
state in $\Lambda $ system takes the form \cite{cptrev}:
\begin{equation}
\left| D\right\rangle =\frac{g_{2}}{g_{0}}\left| 1\right\rangle -\exp \left(
i\Phi \right) \frac{g_{1}}{g_{0}}\left| 2\right\rangle .  \label{D}
\end{equation}
The population of the dark state can be expressed in terms of the
ground-state density matrix elements:
\[
\rho _{DD}=\frac{g_{2}^{2}}{g_{0}^{2}}\rho _{11}+\frac{g_{1}^{2}}{g_{0}^{2}}%
\rho _{22}-\frac{2g_{1}g_{2}}{g_{0}^{2}}%
\mathop{\rm Re}%
\left( \sigma _{21}\exp \left( -i\Phi \right) \right) ,
\]
which are, to the first order in $g_{m}$,
\begin{eqnarray*}
\mathop{\rm Im}%
(\sigma _{21}) &=&\frac{g_{1}g_{2}}{g_{0}^{2}}\sin \Phi -\frac{g_{m}\left(
g_{1}^{2}-g_{2}^{2}\right) }{g_{0}^{4}}, \\
\mathop{\rm Re}%
(\sigma _{21}) &=&-\frac{g_{1}g_{2}}{g_{0}^{2}}\cos \Phi +\frac{g_{m}}{%
g_{0}^{4}}\left( 1+\sin \Phi \right) \cos \Phi , \\
\rho _{11} &=&\frac{g_{2}^{2}}{g_{0}^{2}}-\frac{2g_{m}g_{1}g_{2}}{g_{0}^{2}}%
\sin \Phi , \\
\rho _{22} &=&\frac{g_{1}^{2}}{g_{0}^{2}}+\frac{2g_{m}g_{1}g_{2}}{g_{0}^{2}}%
\sin \Phi .
\end{eqnarray*}
Thus, the population of the dark superposition is $\rho _{DD}=1-\left(
2g_{m}g_{1}g_{2}/g_{0}^{4}\right) \left( 1+\sin \Phi \right) \cos ^{2}\Phi
\approx 1$. It is interesting that a large lower-level coherence is not
established in advance (since $g_{2}(\zeta =0)=0$). However, as soon as $%
g_{2}$ is generated, the coherence emerges, and the medium is prepared in
the nonabsorbing state.

Even in real situation, when both the relaxation rate $\Gamma $ of the
coherence between states $\left| 1\right\rangle $ and $\left| 2\right\rangle
$ and the Doppler broadening are present, the parameters of the process are
in fairly good agreement with calculations presented above. In Fig. 2 the
spatial dependence of the field intensities and the phase $\Phi $ are
plotted for the case when medium is a vapor of Na atoms, excited on D$_{1}$%
-line, at temperature $T=440$ $K$ (this gives the saturated vapor density of
$N=4.42\cdot 10^{11}\,cm^{-3}$ and corresponds to a most probable velocity
of atoms of $v_{p}=5.64\cdot 10^{4}\,cm/sec$), $\Gamma =10^{-4}\gamma $ (1
kHz), input Rabi frequencies $g_{31}\left( \zeta =0\right)
=2.0,\,\,g_{m}=0.02$ (corresponding to intensities of $I_{31}=12.6\,mW/cm^{2}
$ and $I_{m}=1.26\,\mu W/cm^{2}$). We see that dynamics of the intensities
and the phase does not change qualitatively as compared to the case of
negligible decay of the dark state. The behavior of the phase $\Phi $ in
Fig. 2(b) follows the law $\cos \Phi (\zeta )=0$ obtained analytically. The
jumps in the phase occur at points where the intensity of the field being
absorbed approaches zero, according to Eq. (\ref{Maxw5} (b)). The $\omega
_{32}$ wave is generated and reaches its maximum at the length $\zeta =340$
(this corresponds to the real length of the gas cell of $z=1.9\,cm$). This
value is quite close to that calculated from analytical results, Eq. (\ref
{zmax1}), $\zeta _{\max }=314$. The maximum intensity of the $\omega _{32}$
wave is $I_{32}/I_{31}(\zeta =0)=0.952$ which is slightly below the value $%
0.966$ calculated from Eq. (\ref{diss1}) because of an additional
dissipation due to decay of the dark state with the rate $\Gamma $.
Obviously, this rate must be sufficiently small in order to allow for the
population trapping in $\left| D\right\rangle $, namely it must be much
smaller than the optical pumping rate into the dark state:
\begin{equation}
\Gamma /\gamma \ll \frac{g_{0}^{2}}{1+\Delta ^{2}}.  \label{G-cond}
\end{equation}
Here, detuning $\Delta $ includes the Doppler shift: $\gamma \Delta =\Delta
_{31}-k_{31}v_{z}\approx \Delta _{32}-k_{32}v_{z}$, where $\Delta
_{3n}=\omega _{3n}-\left( {\cal E}_{3}-{\cal E}_{n}\right) /\hbar $ are the
laser frequency detunings from transitions $\left| n\right\rangle -\left|
3\right\rangle ,\,(\,n=1,2)$, ${\cal E}_{n}$ is the eigenenergy of the
atomic state $\left| n\right\rangle $. For the resonance $\Delta
_{31}=\Delta _{32}=0$ and large Doppler broadening $k_{31}v_{p}\gg \gamma $,
condition (\ref{G-cond}) reduces to
\begin{equation}
g_{0}^{2}\gg \frac{\Gamma }{\gamma }\left( \frac{k_{31}v_{p}}{\gamma }%
\right) ^{2}.  \label{g0-cond}
\end{equation}
The better this condition is satisfied, the higher the efficiency of
frequency conversion is. The rate $\Gamma $ is determined by m.w. field
fluctuations, atomic collisions and other random phase disturbing processes.
For parameters of the proposed here experiment, $\Gamma $ can be very small.
Recently, the rate $\Gamma <50\,Hz$ has been observed in experiment \cite
{wyn97}.

Inasmuch as CPT is a basis for the considered above scheme, the generation
occurs under quite restrictive conditions on e.m. wave frequencies. It is
well known that CPT takes place when the optical frequencies are in the
narrow range (''black line'') around two-photon resonance \cite{cptrev}:
\begin{equation}
\Delta _{32}=\Delta _{31}.  \label{2ph-cond}
\end{equation}
Considering that the wave $E_{2}$ is always generated at frequency $\omega
_{32}=\omega _{31}-\omega _{m}$ (simply due to photon energy conservation),
the condition (\ref{2ph-cond}) tells us that the generation takes place only
if the m.w. field frequency is in the narrow range around resonance with
transition $\left| 1\right\rangle -\left| 2\right\rangle $: $\omega
_{m}=\left( {\cal E}_{2}-{\cal E}_{1}\right) /\hbar $. Figure 3 demonstrates
this fact. The width of the generation peak (width of the black line) is
determined by the pumping rate into the dark state \cite{cptrev}. In the
presence of large Doppler broadening this width is of the order of $\delta
\omega _{m}\approx $ $g_{0}^{2}/\left( k_{31}v_{p}/\gamma \right) ^{2}\gamma
$, which is a few kHz for parameters of Fig. 3.

It is interesting that, at the same time, the large Doppler broadening
allows for a broad tuning of the generated wave. In Fig. 4 we have plotted
dependence of the generated intensity on detuning $\Delta _{31}$ (for fixed $%
\omega _{m}=\left( {\cal E}_{2}-{\cal E}_{1}\right) /\hbar $) at the optical
length $\zeta =340$. One can see that conversion efficiency remains fairly
large for detunings of the order of the Doppler broadening (few GHz for Na
vapor). This is because the CPT survives even at large common detunings $%
\Delta _{32}=\Delta _{31}$ as long as the condition (\ref{G-cond}) is
satisfied.

The second mechanism of EIT allowing efficient frequency conversion in the $%
\Lambda $ medium takes place at strong m.w. fields, $g_{m}\gg 1,\,g_{0}$. In
this case, the absorption coefficients to the second order in $\left(
1/g_{m}\right) $ are:
\begin{mathletters}
\label{ims-2}
\begin{eqnarray}
\mathop{\rm Im}%
(\sigma _{31}) &=&\frac{g_{2}}{2g_{m}}\sin \Phi +\frac{g_{1}\left(
g_{1}^{2}-g_{2}^{2}+2g_{2}^{2}\sin ^{2}\Phi \right) }{2g_{m}^{2}g_{0}^{2}},
\\
\mathop{\rm Im}%
(\sigma _{32}) &=&-\frac{g_{1}}{2g_{m}}\sin \Phi -\frac{g_{2}\left(
g_{1}^{2}-g_{2}^{2}-2g_{1}^{2}\sin ^{2}\Phi \right) }{2g_{m}^{2}g_{0}^{2}}.
\end{eqnarray}
The energy dissipation is determined by the equation
\end{mathletters}
\[
\frac{dg_{0}^{2}}{d\zeta }=-\frac{g_{0}^{2}}{g_{m}^{2}}\,,
\]
with a solution
\begin{equation}
g_{0}^{2}=g_{0}^{2}(\zeta =0)\exp \left( -g_{m}^{-2}\zeta \right) .
\label{diss2}
\end{equation}
Again, if we neglect the slow total energy dissipation (i.e., we neglect
terms of the second order in $g_{m}$ in Eq. (\ref{ims-2})), we obtain the
solution of amplitude equations, very similar to the CPT case:
\begin{mathletters}
\label{sol2}
\begin{eqnarray}
g_{1}^{2} &=&g_{0}^{2}\cos ^{2}\left( \frac{1}{2g_{m}}\zeta \right) , \\
g_{2}^{2} &=&g_{0}^{2}\sin ^{2}\left( \frac{1}{2g_{m}}\zeta \right) .
\end{eqnarray}
Here, the period of intensity oscillations is $\zeta _{\pi }=2\pi g_{m}$,
which is again much smaller than the characteristic length of the total
energy dissipation: $\zeta _{diss}\approx g_{m}^{2}$. Maximum energy
transfer to the $\omega _{32}$ field occurs at
\end{mathletters}
\begin{equation}
\zeta _{\max }=\pi g_{m}.  \label{zmax2}
\end{equation}
The loss of the optical intensity is $\Delta I\,/\,I=1-\exp \left( -\pi
/g_{m}\right) \ll 1$ at this point.

Numerical calculations of the optical waves propagation give the results
which are in very good agreement with analytical ones, and which are
qualitatively very similar to the CPT-case in Fig. 2. The physical mechanism
is, however, different. As we have discussed above, CPT does not work at
strong m.w. fields except under the specific conditions (\ref{phi-cond}) and
(\ref{g-cond}). This can be proved by considering the density matrix
elements. It turns out that for this case the ground state populations are
equal to $\rho _{11}=\rho _{22}=0.5$ up to the second order in $\left(
1/g_{m}\right) $, $%
\mathop{\rm Im}%
(\sigma _{21})=O\left( 1/g_{m}^{2}\right) $ and $%
\mathop{\rm Re}%
(\sigma _{21})=-\left( g_{1}g_{2}/g_{0}^{2}\right) \cos \Phi +O\left(
1/g_{m}^{2}\right) $. Therefore, the atomic population is, for arbitrary $%
g_{1},g_{2}$ and $\Phi $, not all pumped into the dark state: $\rho
_{DD}=1/2+\left( 2g_{1}^{2}g_{2}^{2}/g_{0}^{4}\right) \cos ^{2}\Phi <1$.
Under the condition (\ref{cos0}) taking place at the generation, the
ground-state coherence $\sigma _{21}$ is negligibly small and $\rho _{DD}=1/2
$.

Thus, a very weak optical absorption at strong m.w. field is due to the
Autler-Townes effect, or, in other terms, due to the capture of almost all
atomic population by strong m.w. field in two-level system $\left|
1\right\rangle -\left| 2\right\rangle $. This mechanism requires large m.w.
intensities, but it has some advantages over the CPT-case. First of all, it
is more robust. For example, the relaxation $\Gamma $ does not play so
important role, and the generation range of m.w. frequency $\omega _{m}$ is
much broader (it is of order of $g_{m}$) as compared to the case of weak
m.w. field. Similar to the case with a weak m.w. field, there is a
possibility to tune the generated radiation over the Doppler contour, and
here the tuning is not as sensitive to the value of $\Gamma $ as in former
case. Note that here the optical length scales are determined only by $g_{m}$
and do not depend on the input intensity, $g_{0}^{2}(\zeta =0)$. Therefore,
the maximum conversion takes place at the same length for different input
intensities. This is especially important for experiments with nonuniform
light beams, e.g., with the Gaussian intensity profile. Another important
advantage is that this mechanism can be applied not only to $\Lambda $%
-system, but also to a V-scheme with one ground and two excited states. We
believe that the optical frequency conversion may be observed
experimentally, for example, in a V-system of $Pr^{3+}:YAlO_{3}$ solid where
the reduced absorption was recently demonstrated \cite{yam98}.

Finally, Fig. 5 represents the dependence of generated wave on Rabi
frequency $g_{m}$ of the m.w. field at fixed optical length $\zeta =340$.
This figure clearly demonstrates two ranges of $g_{m}$ where EIT and,
correspondingly, efficient frequency conversion occur.

In summary, we have described a scheme for efficient optical conversion
based on EIT in atomic $\Lambda $ system where the interaction loop is
closed by a microwave field. Depending on the m.w. intensity, two mechanisms
of EIT work in this scheme: CPT and Autler-Townes effects. Intensity of the
m.w. field plays also a role of controlling parameter in the conversion
process - it determines optical length scales of the process, cf. Eqs. (\ref
{zmax1}), (\ref{zmax2}), as well as the degree of the total energy
dissipation, cf. Eqs. (\ref{diss1}), (\ref{diss2}). An optimal choice is
always possible, which should allow an experimental realization of the
proposed scheme in different systems. Since the EIT-assisted frequency
conversion combines large nonlinearity with substantially reduced
spontaneous emission noise, one may expect that the generated signal will be
fluctuation-correlated with the pump wave \cite{hemm97}. Such correlations
persist even on a quantum level \cite{luk99}. Therefore, the present scheme
can be used for generation of two phase-correlated optical waves, which
would be an alternative to conventional methods using electro- or
acousto-optical modulators, or direct current modulation in laser diodes.
Another possible application may be a generation of squeezed light \cite
{kol93}.

\section{Acknowledgments}

We are very grateful to Prof. L. Windholz for his continuous interest to
this work and useful discussions. D.V. Kosachiov thanks the members of the
Institut f\"{u}r Experimentalphysik, TU Graz, for hospitality and support.
This study was supported by the Austrian Science Foundation under project
No. P 12894-PHY.

\newpage

{\bf \centerline{\Large{\bf Figure captions}}}

\bigskip

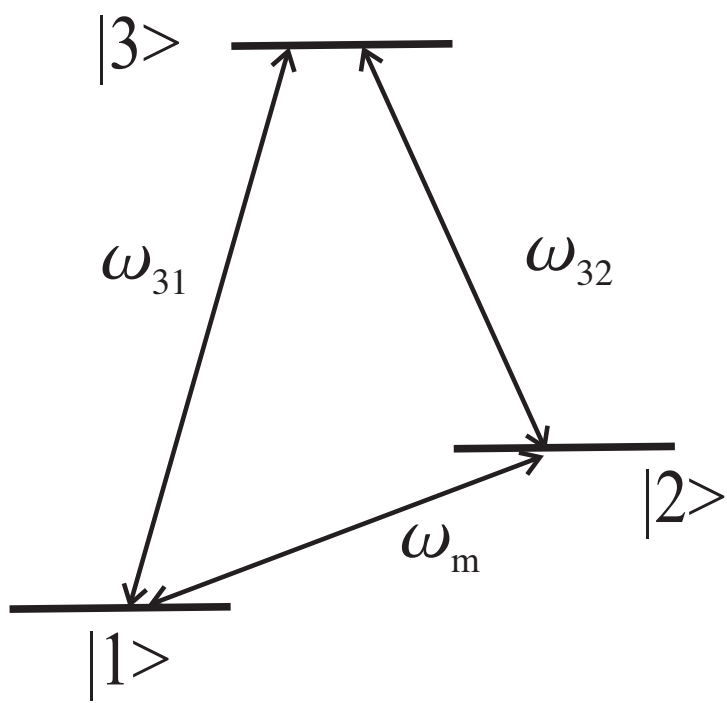
Fig. 1. $\Lambda $ system with two metastable states $\left| 1\right\rangle $
and $\left| 2\right\rangle $. $\omega _{31}$ and $\omega _{32}$ are the
optical frequencies, $\omega _{m}$ is the microwave frequency.

\medskip

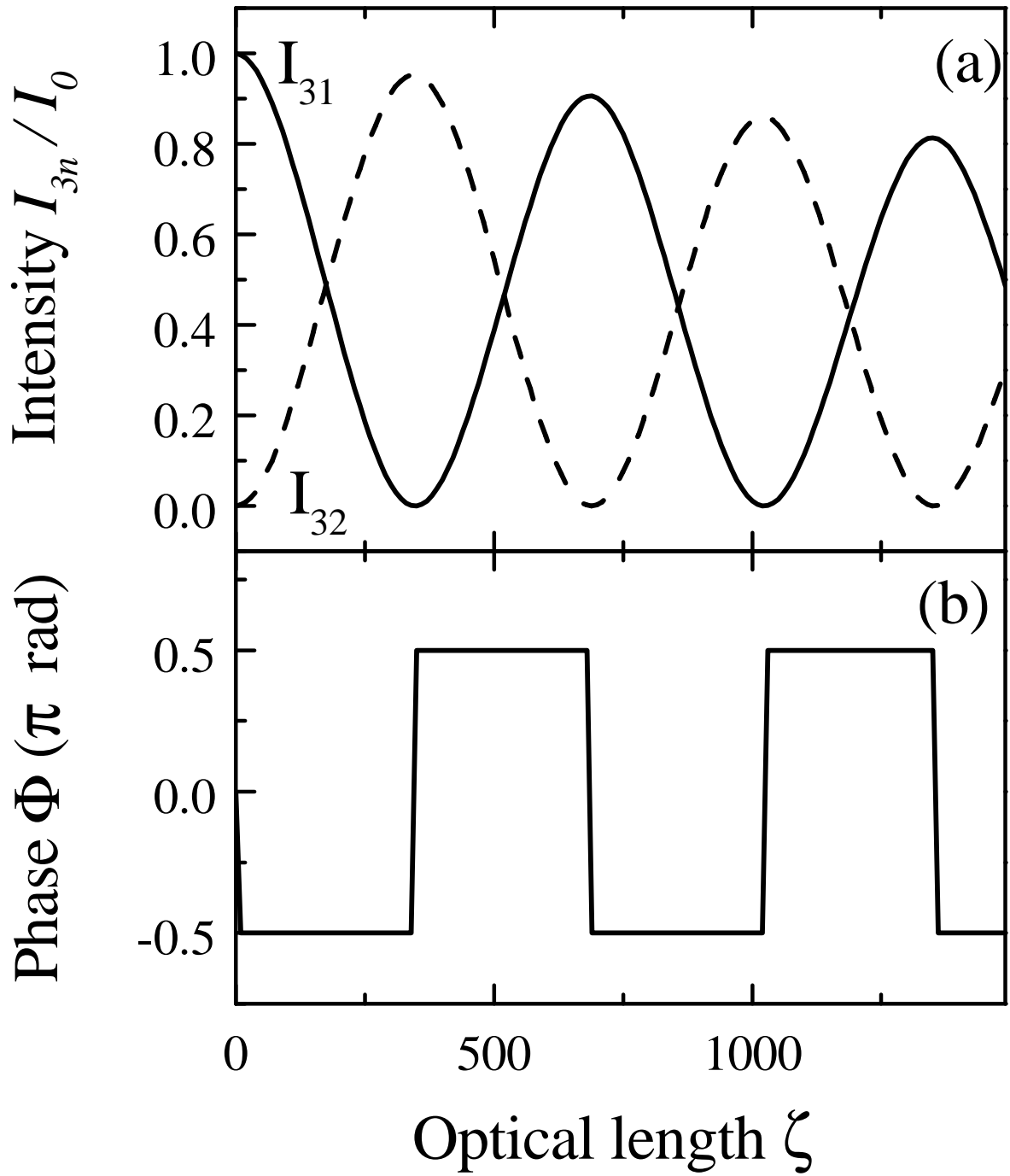
Fig. 2. Spatial variations of: (a) optical field intensities $I_{31}$ (solid
curve) and $I_{32}$ (dashed curve) in units of input intensity $I_{0}\equiv
\,I_{31}\left( \zeta =0\right) $, (b) the relative phase $\Phi $, in a vapor
of $^{23}$Na atoms interacting with radiation in a $\Lambda $ configuration
of levels $3^{2}S_{1/2}(F=1)-3\,^{2}S_{1/2}(F=2)-$ $\,3^{2}P_{1/2}$. Vapor
temperature $T=440$ $K$, $\Gamma =10^{-4}\gamma $, detunings $\Delta
_{31}=\Delta _{32}=0$, Rabi frequencies of input fields $g_{31}\left( \zeta
=0\right) =2.0,\,\,g_{m}=0.02$.

\medskip

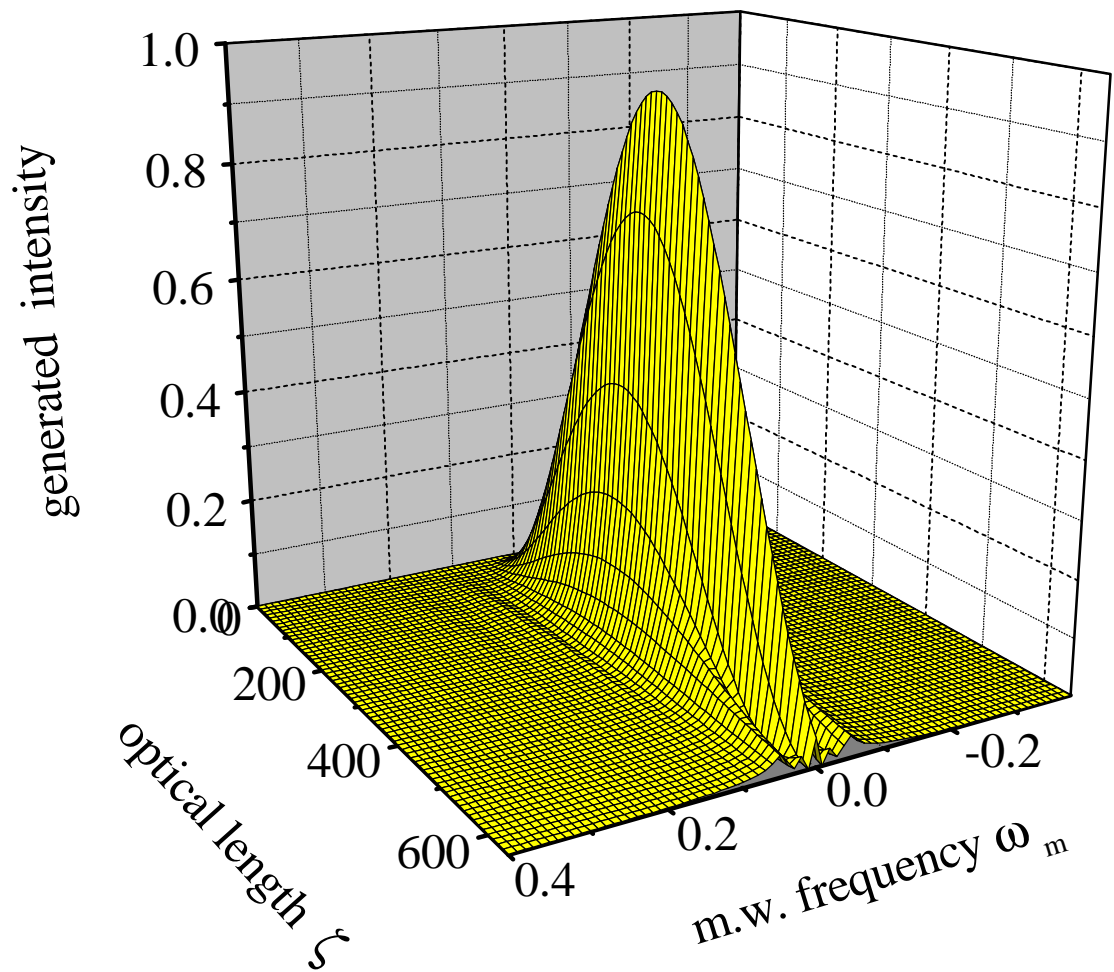
Fig. 3. Generation of the $E_{2}$ wave (in units of input intensity $%
I_{0}\equiv \,I_{31}\left( \zeta =0\right) $) as a function of the microwave
frequency $\omega _{m}$ (in units of the excited state relaxation rate $%
\gamma $). Other parameters are the same as in Fig. 2.

\medskip

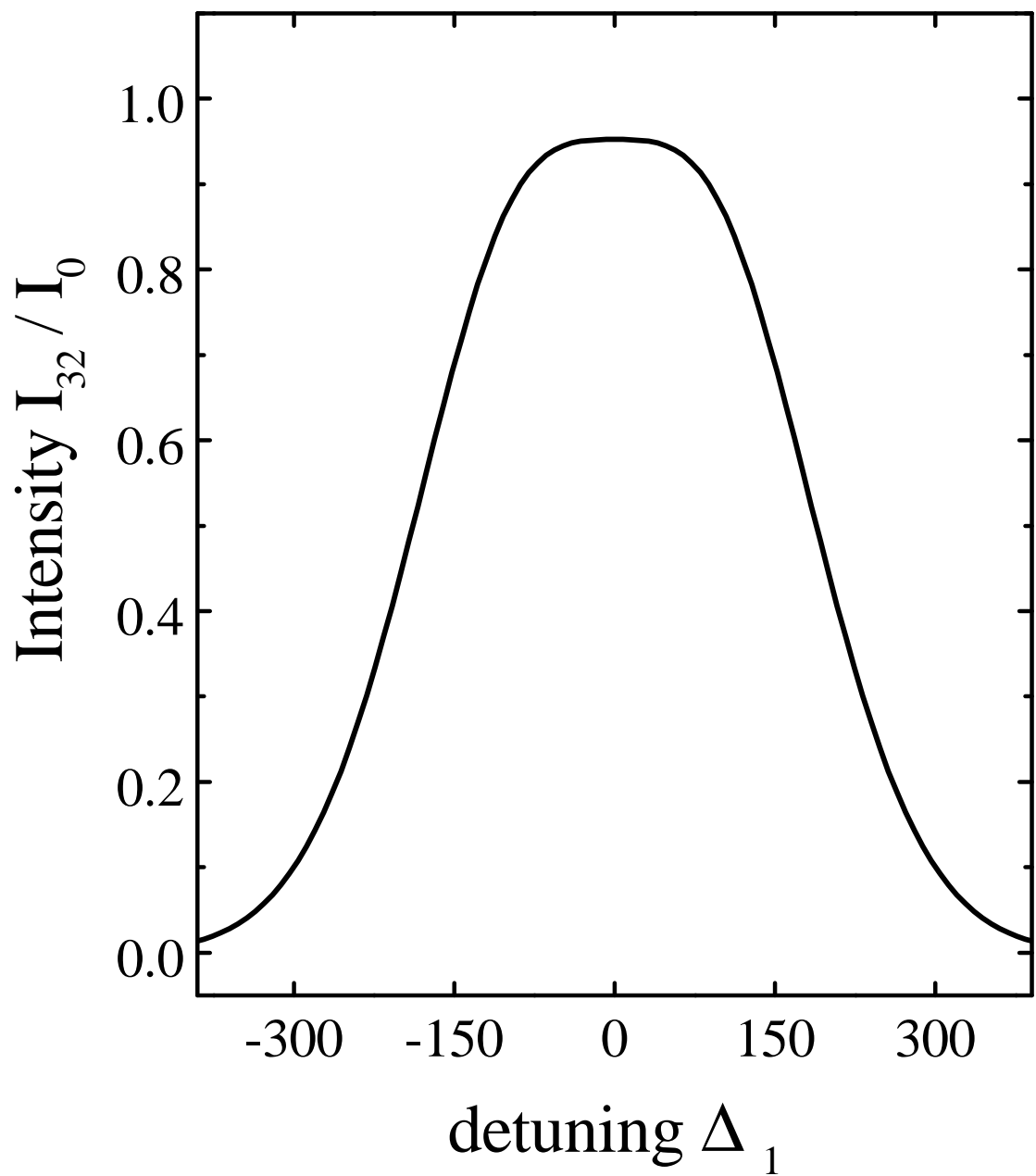
Fig. 4. Dependence of the generated intensity $I_{32}/I_{0}$ on detuning $%
\Delta _{31}$ (in units of the excited state relaxation rate $\gamma $) for
fixed $\omega _{m}=\left( {\cal E}_{2}-{\cal E}_{1}\right) /\hbar $, at the
optical length $\zeta =340$. Other parameters are the same as in Fig. 2.

\medskip

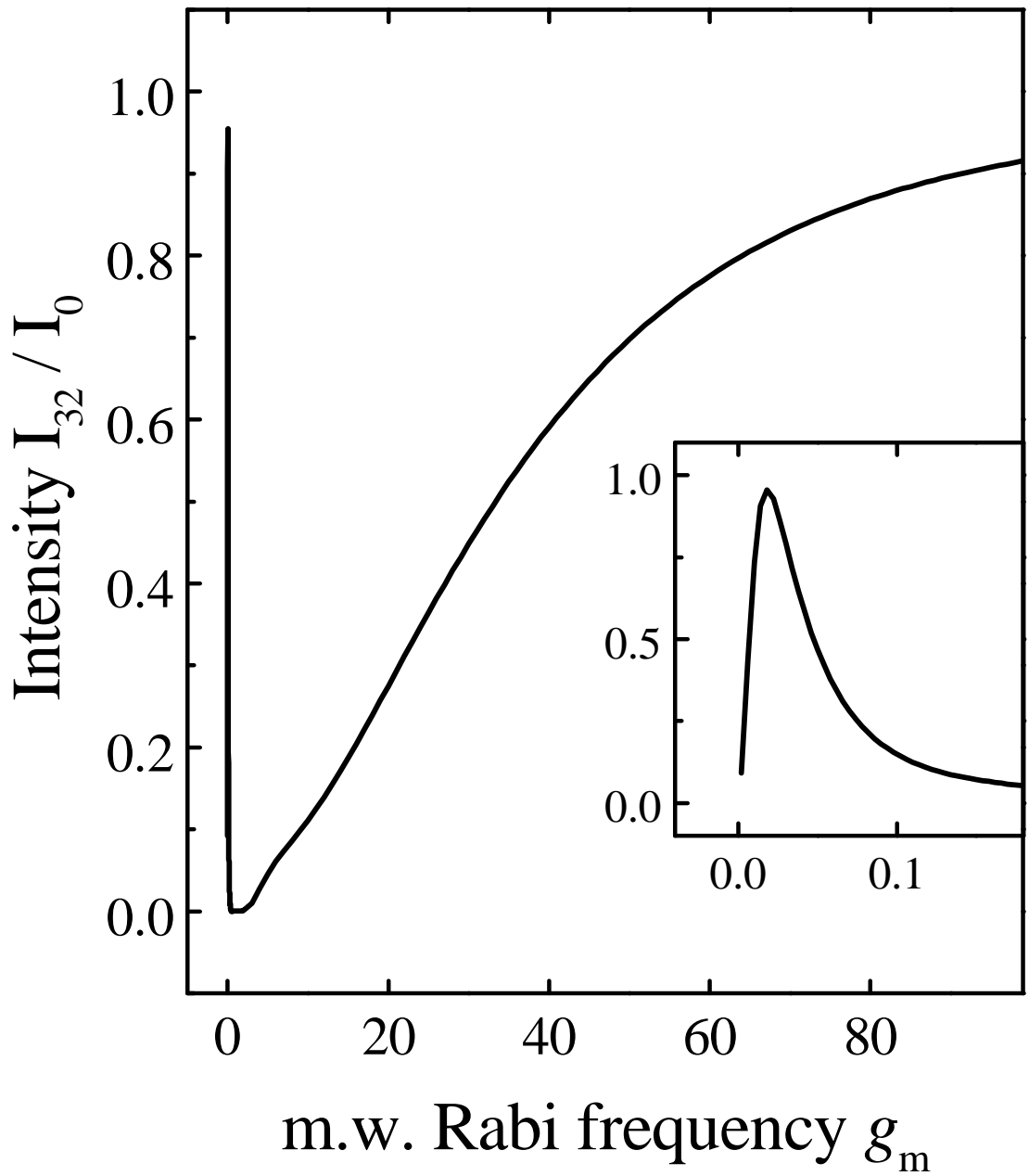
Fig. 5. Dependence of the generated intensity $I_{32}/I_{0}$ on the Rabi
frequency of microwave field $g_{m}$ at the optical length $\zeta =340$.
Other parameters are the same as in Fig. 2. Inset shows the range of small $%
g_{m}$.

\end{document}